\magnification= 1200
\baselineskip 18pt

\def \br{{\bf r}}

\centerline{\bf ON EDGE STATES IN SEMI-INFINITE QUANTUM HALL SYSTEMS}

\vskip 0.5cm

\centerline{N. MACRIS, P. A. MARTIN}
\vskip 0.25cm
\centerline{Institut de Physique Theorique}
\centerline{Ecole Polytechnique Federale de Lausanne}
\centerline{CH - 1015 Lausanne}

\vskip 0.5cm

\centerline{J. V. PUL\'E}
\vskip 0.25cm
\centerline{Department of mathematical Physics}
\centerline{University College Dublin}
\centerline{Belfield 4, Dublin}

\vskip 0.5cm

\centerline{\bf Abstract}

\noindent We consider an electron in two dimensions submitted to a magnetic 
field and to the potential of impurities. We show that when the electron is confined
to a half-space by a planar wall described by a smooth increasing potential, 
the total Hamiltonian necessarily has a continuous spectrum in some intervals 
in-between the Landau levels provided that both the amplitude and spatial 
variation of the impurity potential are sufficiently weak. 
The spatial decay of the 
impurity potential is not needed. In particular this proves the occurence of 
edge states in semi-infinite quantum Hall systems.

\vfill

\beginsection{1. INTRODUCTION}

In the quantum Hall effect the physics at the boundary of the sample 
plays an important role. This was first recognized in the context of the integer
effect by Halperin [1], and has been later on the subject of powerful
and beautiful theories of the fractional effect [2,3,4]. Moreover the edge properties of
quantum Hall fluids are now experimentally accessible [5] and provide much 
information on these systems. Halperin argued that in a two dimensional
system with boundaries (say an annulus or a cylinder) submitted to a perpendicular
magnetic field there are "quasi one-dimensional edge states" extended along 
each boundary of the sample, which contribute to the quantized Hall 
conductivity (if the chemical potentials on opposite edges are different). In the 
ideal situation of a clean sample with non-interacting electrons it is easy to 
construct such quasi one-dimensional edge states (see (1.6) below). In a real situation
however, the sample is disordered (we disregard Coulomb interactions here) and the 
existence of extended states may seem less evident since usual one-dimensional
systems exhibit localization for any amount of disorder. In fact the absence of
localization in this situation is usually explained by the chirality of the modes 
propagating along the boundary. The simplest models for investigating the 
boundary of quantum Hall fluids
are  constructed from one-dimensional electron systems with chiral 
branches of exitations (see for example [2,4]) corresponding
to only left movers (or only right movers). In such systems the absence of interference 
between left and right movers leads to the absence of Anderson localization
and hence to extended states.

In the present work we consider a two-dimensional semi-infinite geometry, the electrons 
being confined by a planar smooth wall. Starting right away from an unbounded system
allows a clearcut distinction between point and continuous spectrum in the 
Schroedinger operator. We show here that for a large class of additional impurity
potentials (random or deterministic) the Hamiltonian of this system has indeed 
continuous parts in its spectrum as a result of the presence of the wall. We recall that for 
the case of infinite space (system without boundary) and point random impurities
the spectrum has been shown to be {\it entirely dense pure point} for large 
enough magnetic field [6], 
a fact which is believed to hold for more general random potentials
\footnote*{This is compatible with a
localization length which may diverge at isolated energies.}. For these, 
 existing results deal only with limited regions of the spectrum [7,8,9].
Here we are mainly concerned with the opposite problem, namely the exclusion of point spectrum
in certain parts of the spectrum\footnote*{We do not exclude the possibility of a singular continuous
spectrum.}.
During the completion of this work, we have been informed that Froehlich, 
Graf and Walcher [10] have proven the existence of intervals of absolute 
continuity for the same 
system with the help of Mourre estimates, and have also treated the case of Dirichlet boundary
conditions. 

The position of the electron in the plane is $\br=(x, y)$, $y$ 
being the direction 
parallel to the wall. We assume that the  
"wall potential" $U(x)$ vanishes for $x\leq 0$ (the bulk region) 
and increases as a power
law for $x>0$ (the wall region)
$$
U(x)=0,\qquad x\leq 0, \qquad U(x)=\mu x^\gamma,\qquad x\geq 0
\eqno(1.1)
$$ 
where $\mu>0$ and $\gamma\geq 1$.
All our  estimates hold for  wall potentials that grow much faster as long as they 
are continuous on the whole real line
and strictly
increasing (for example exponential or gaussian growth) but we limit ourselves
to (1.1) for the sake of concreteness.

In addition
the electron is also submitted to a bounded and differentiable
 external potential $w(\br)$ such that
$$
\sup_{\br}\vert w(\br)\vert =w_0<\infty, \qquad\qquad 
\sup_{\br}\vert \partial_x w(\br)\vert =w_0^\prime<\infty
\eqno(1.2)
$$
The total Hamiltonian  is
$$
H=H_0+w(\br)
\eqno(1.3)
$$
where $H_0$ is the pure wall Hamiltonian
$$
H_0=H(B)+U(x)
\eqno(1.4)
$$
and 
$$
H(B)={1\over 2}p_x^2+{1\over 2}(p_y-Bx)^2
\eqno(1.5)
$$
is the usual Landau Hamiltonain written in the Landau gauge, $B$ being the strength of the magnetic field.
All these Hamiltonians are essentially self-adjoint on $C_0^\infty({\bf R^2})$ [11, theorem X34, p.190].

It is well known that the spectrum of $H(B)$ consists of Landau levels 
$E_n=(n+{1\over 2})B$, $n=0,1,2...$. These are infinitely degenerate  so that one
may construct localized as well
as extended eigenfunctions of $H(B)$ by appropriate linear combinations. Because of  
translation invariance in the $y$ direction the generalized eigenfunctions
of $H_0$ are of the form
$$
e^{iky} h_{nk}(x)
\eqno(1.6)
$$
where $h_{nk}(x)$ is the solution of the one dimensional  problem 
${\cal H}(k)h_{nk}={\cal E}_n(k)h_{nk}$,
$$
{\cal H}(k)={1\over 2}p_x^2+{1\over 2}(k-Bx)^2+U(x)
\eqno(1.7)
$$
For a given $k$ this is the Hamiltonian of a particle in a confining potential
well both for $x\to\pm \infty$. Therefore the spectrum consists of 
non degenerate eigenvalues ${\cal E}_n(k)$ and it follows from the Kato-Rellich theorem
[12] that the branches 
${\cal E}_n(k)$ are analytic functions of $k$.

Note also that from the  Feynman-Hellman theorem
$$
\eqalignno{\partial_k{\cal E}_n(k)&=
\int_{-\infty}^{+\infty} dx (k-Bx)\vert h_{nk}(x)\vert^2&(1.8a)\cr
&={1\over B}\int_{-\infty}^{+\infty} dx U^\prime(x)\vert h_{nk}(x)\vert^2&(1.8b)\cr}
$$
Since $U^\prime(x)>0$ for $x>0$ the branches are strictly monotone
increasing for all $k$. The equality (1.8b) follows from the fact that the Hamiltonian 
(1.7) is unitarily equivalent to 
$$
{1\over 2}p_x^2+{1\over 2}B^2x^2+U(x+{k\over B})
\eqno(1.9)
$$
Moreover, $\lim_{k\to -\infty}{\cal E}_n(k)=(n+{1\over 2})B$ since $U(x+{k\over B})$
vanishes in this limit, whereas  
$\lim_{k\to +\infty}{\cal E}_n(k)=+\infty$ since $U(x+{k\over B})$ is unbounded 
as $k\to+\infty$. One concludes from these observations that
the spectrum of the wall Hamiltonian  $H_0$ 
is absolutely continuous and consists of the set
$\sigma(H_0)=[{B\over 2}, +\infty[$.  

The right hand side of (1.8a) is the average of the diamagnetic current $j_y=p_y-Bx$
carried by the state (1.6) 
along the wall and $\partial_k{\cal E}_n(k)$ can be 
interpreted as the corresponding group velocity.

The main question that we address in this paper is the stability of this continuous 
spectrum (or parts of it) when an impurity potential $w$ is added to $H_0$.
 Consider first the case when $w$ is localized in a finite region of space. Then, as 
shown in Appendix A, $e^{-tH}-e^{-tH_0}$ is a trace class 
operator, implying that $H$ still has absolutely continuous
spectrum in $[{B\over 2}, +\infty[$ [13, chap X]. If $w$ does not decay at infinity, as 
is the case for example for  periodic or  random potentials, this argument 
cannot be applied and the nature of the spectrum may be different. 
In the random case a convenient model of disorder
is obtained by assuming that the impurities are located
on a regular lattice, each of them being the source of the same local
bounded potential, with bounded derivatives. Then for $w(\br)$ we take 
 a typical realization of the random potential
$$
V_\omega(\br)=\sum_{(n,m)\in {\bf Z^2}} \omega_{n,m}v(x-n, y-m),\qquad
v(\br)=0\qquad{\rm for }\qquad \vert\br\vert\geq {1\over 2}
\eqno(1.10)
$$
where $\omega_{n,m}$ are independent identically distributed
random variables with continuous density supported in $[-1,1]$.
If $v(\br)$ satisfies (1.2) the same is true for all $V_\omega(\br)$.
Typical realizations of the random potential
do not decay at large distance, so that trace class perturbation
theorems cannot be applied. 

We prove in section 4, theorem 1, that for $w_0$ and $w_0^\prime$ small enough,
depending on $B$ and the steepness of the wall, $H$ cannot 
have a point spectrum in intervals $\Delta_n(B,\delta)=](n+1)B-\delta, (n+1)B+\delta[$
of size $2\delta>B-w_0$ in-between the Landau levels. This information is supplemented 
in section 5, theorem 2, by a proof that the intervals $\Delta_n(B,\delta)$ are 
included in the spectrum of $H$. Hence theorems 1 and 2 imply that
the spectrum of $H$ remains continuous in these
intervals. In the random case, general arguments (Appendix B) show that the spectrum of $H_\omega$ contains $[{B\over 2},\infty[$ with probability one, and 
thus theorem 1 alone implies the existence of continuous spectrum for almost
all $\omega$. Theorem 2 guarantees the existence of a continuous spectrum
for all realizations of (1.10) and allows us to treat deterministic potentials
(e.g periodic).

\beginsection{2. MAIN RESULTS}

\noindent In the subsequent sections, the effect of the wall will 
 enter mainly through 
 the following 
functional of the wall potential
$$
A(E; U)=\sup_{0\leq x\leq x_0(U)}\biggl({U(x)^4\over U^\prime(x)}\biggr)
+8\int_{x_0(U)}^\infty dx {U(x)^4 U({x\over 2})^{1/4}\over 
(2\pi x)^{1/2}U^\prime(x)}
\exp\bigl(-{x\over 2\sqrt 2} U({x\over 2})^{1/2}\bigr)
\eqno(2.1)
$$
where $E> 0$ and $x_0(U)$  solves $U(x_0/2)=2E$. We remark that the 
 $A(E;U)$ is finite for a much larger class of wall potentials
than those in (1.1) (e.g exponential). 
The factor
$$
\exp\bigl(-{x\over 2\sqrt 2} U({x\over 2})^{1/2}\bigr)
$$
is similar to the  WKB tail
of the wavefunction $\exp(-\int^xdy\sqrt{U(y)-E})
\sim \exp(-cx\sqrt{U(x)})$ in the wall region.
We have
\vskip 0.5cm

\noindent {\bf Theorem 1}

\noindent {\it If ${B\over 2}-w_0>\delta$ for some $\delta>0$ and 
$$
w_0^\prime<
{({B\over 2}-\delta-w_0)^4\over \sup_{E\in \Delta_n(B,\delta)} A(E+w_0;U)}
\eqno(2.2)
$$
then $H$ has no eigenvalue in the 
intervals $\Delta_n(B,\delta)=](n+1)B-\delta, (n+1)B+\delta[$.}

\vskip 0.5cm

\noindent{\bf Theorem 2}

\noindent{\it If ${B\over 2}-w_0>\delta$ for some 
$\delta>0$ and 
$$
w_0^\prime<
{({B\over 2}-\delta-w_0)^4\over
\sup_{0\leq w_0\leq{B\over 2}} \sup_{E\in \Delta_n(B,\delta)} A(E+w_0;U)}
\eqno(2.3)
$$
then 
the whole interval $\Delta_n(B,\delta)$ is included in the spectrum of H.}
 
\vskip 0.5cm

Note that condition (2.3) implies (2.2), hence under condition (2.3)
the spectrum of  $H$  in $\Delta_n(B,\delta)$ is purely 
continuous.

In order to understand the meaning of this condition more explicitely, let us consider
for concreteness the case of a "linear wall", $U(x)=\mu x$, $x\geq 0$ and 
$U(x)=0$, $x\leq 0$ with $\mu>0$. Then we have
$$
A_{\rm linear}(E)={(4E)^4\over \mu}+{8\over \sqrt{2\pi} 2^{1/4}}
\mu^{5/3}
\int_{{4E\over \mu^{2/3}}}^\infty dx x^{15/4} 
e^{-{1\over 4} x^{3/2}}
\eqno(2.4)
$$
Consider first the dependence on the steepness $\mu$ of the 
wall for fixed $B$ and $n$.
We find
$$
w_0^\prime< Cst.\mu,\qquad \mu\to 0
\eqno(2.5)
$$
On the other hand for a steep wall, $\mu\to \infty$, we find
the condition
$$
w_0^\prime< Cst. \mu^{-5/3}, \qquad \mu\to +\infty
\eqno(2.6)
$$
The case (2.5) shows that our theorems can only prove
the occurence of a continuous spectrum
 if the wall is steep enough. For a localized impurity ($w(\br)$ with
compact support) we know that the spectrum of $H$ has  a continuous part
 for any $\mu>0$
(see introduction). When the impurities are extended over the whole space
it is not known whether the same is true, or if 
 a critical steepness is needed for the occurence of a continuous spectrum. 
On the other hand, because of (2.6) we cannot conclude about 
the existence of a continuous spectrum
for a very steep wall. We nevertheless believe that the system still has extended
edge states in the limit of hard walls. The use of non-linear walls in (2.1)
(e.g polynomial, exponential,...) leads to the same conclusion that an upper
bound on $w_0^\prime$ has to vanish in the limit of 
infinite steepness. We feel that this inability to prove the existence of edge states 
for very steep walls is linked to the techniques used in this paper.

We consider now $B$ large and $\mu$, $n$ fixed. Then one can allow $w_0$ to be large
but
$w_0^\prime$ has to remain bounded because it follows from (2.3) 
and (2.4) that $w_0^\prime\leq C$, $C> 0$
 as $B\to\infty$. However for a non linear wall 
 $U(x)=\mu x^\gamma$, $x>0$ with $\gamma>1$ 
the latter bound can be improved to $w_0^\prime\leq CB^{1-{1\over \gamma}}$
allowing for more rapidly varying impurity potentials.

Finally the high energy (i.e large $n$) behaviour of the bound (2.3) is as $n^{-4}$.
Thus theorems 1 and 2 guarantee the existence of continuous spectrum only
in a finite number of intervals. We expect however to find
a continuous spectrum
 in-between all Landau levels, especially at high energies. This is probably
again an artefact of our methods.

To conclude this section,
we say a few words about the ideas involved in the proof of Theorems 1 and 2.
The absolute continuity of $\sigma(H_0)$ is intimately linked to the positive 
group velocity $\partial_k {\cal E}_n(k)$ and positive "local drift" $U^\prime(x)/B$
in (1.8b). A closely related fact is the existence of a positive commutator of $H_0$
with the observable 
$$
Y=p_x-By
$$ 
which represents the centre of the cyclotronic orbit
for the Landau problem
$$
[iY,H_0]=U^\prime(x)\geq 0
\eqno(2.7)
$$
Suppose there  exist an eigenfunction
$\Psi_0$ of $H_0$. Then
$$
0=\langle\Psi_0, [iY,H_0]\Psi_0\rangle=\int_0^{+\infty}dx
\int_{-\infty}^{+\infty}dy
U^\prime(x)\vert\Psi_0(x,y)\vert^2
\eqno(2.8)
$$
which is clearly impossible because $\Psi_0$ has a tail penetrating the region
$x>0$ where $U^\prime(x)$ is strictly positive. This observation can be 
generalized to the full Hamiltonian (2.3). The commutator with $Y=p_x-By$ becomes
$$
[iY,H]=U^\prime(x) + \partial_x w(\br)
\eqno(2.9)
$$
Now we loose the positivity because of $\partial_x w$, but we may exploit the fact that $\partial_x w$ is bounded and 
$U^\prime(x)\to +\infty$ for $x\to +\infty$. If $H$ has an eigenstate
$\Psi$, then $\langle\Psi, [iY,H]\Psi\rangle =0$ so that
$$
\langle \Psi,U^\prime\Psi\rangle=-\langle \Psi,\partial_x w\Psi\rangle\leq w_0^\prime
\eqno(2.10)
$$
Suppose now that the eigenenergy of $\Psi$ lies in-between two Landau levels,
then $\Psi$ should be supported in regions where $U(x)$ is large. 
Indeed, if $\Psi$ were essentially localized in the bulk region, the wall
would not contribute to the energy which would then lie in the vicinity of
a Landau level for small $w_0$.
Therefore
$\langle \Psi,U^\prime\Psi\rangle$ should be large, which contradicts (2.10) if 
$w_0^\prime$ is sufficiently small.

The basic idea 
behind the proof of theorem 2 is as follows. Suppose that 
$E\in \Delta_n(B,\delta)$ belongs to the resolvent set of $H$. Then there is
a small gap around $E$ in the spectrum of $H$. It is possible to add a local
perturbation to $H$ such that: i) theorem 1 is still applicable to the 
perturbed Hamiltonian, ii) the local perturbation  creates eigenvalues 
inside the gap. From i) and ii) we obtain a contradiction so that 
$E\in \Delta_n(B,\delta)$ cannot belong to the resolvent set of $H$.

\beginsection{3. POINT-WISE ESTIMATE OF WAVEFUNCTIONS }

In this section we provide a control on the decay of the eigenfunctions
of $H$ for $x>0$ using Brownian motion techniques. Let ${\bf b}(s)=(b_x(s),b_y(s))$ be a
two-dimensional Brownian path with $0\leq s\leq t$ and ${\bf b}(0)=0$. We denote
by $D{\bf b}=Db_xDb_y$ the Wiener measure with
covariance $\langle b_i(s) b_j(s)\rangle=\min(s,t)\delta_{ij}$, $i,j=x,y$.
For an eigenfunction $\Psi$ with eigenvalue $E$ we have, 
using the Feynman-Kac-Ito representation [14],
$$
e^{-tE}\Psi(\br)=\int D{\bf b} e^{-iB\int_0^t (x+ b_x(s))db_y}
e^{-\int_0^tds \bigl(U(x+b_x(s))+w(\br+{\bf b}(s))\bigr)}\Psi(\br+{\bf b}(t))
\eqno(3.1)
$$
Using the Schwarz inequality on the measure $D{\bf b}$, hypothesis (1.2), and
then integrating over the $y$ 
direction, we get
$$
e^{-2tE}\int_{-\infty}^{+\infty} dy\vert\Psi(\br)\vert^2
\leq 
e^{2tw_0}
\int D b_x 
e^{-2\int_0^tds U(x+b_x(s))}\int D b_x F(x+b_x(t))
\eqno(3.2)
$$
where 
$$
F(x+b_x(t))=\int_{-\infty}^{+\infty} dy\vert\Psi(x+b_x(t),y)\vert^2
\eqno(3.3)
$$
We have
$$
\eqalign{\int D b_x F(x+b_x(t))&=\int_{-\infty}^{+\infty} dx^\prime 
{e^{-{x^{\prime 2}\over 2t}}\over (2\pi t)^{1/2}} F(x+x^\prime)
\leq {1\over (2\pi t)^{1/2}}\int_{-\infty}^{+\infty} dx^\prime F(x^\prime)
\cr &
={1\over (2\pi t)^{1/2}}\int d\br \vert\Psi(\br)\vert^2={1\over (2\pi t)^{1/2}}
\cr}
\eqno(3.4)
$$
From (3.2) and (3.4) we have the estimate
$$
\int_{-\infty}^{+\infty} dy\vert\Psi(\br)\vert^2
\leq 
{e^{2t(E+w_0)}\over (2\pi t)^{1/2}}
\int Db_x 
e^{-2\int_0^tds U(x+b_x(s))}
\eqno(3.5)
$$
Set now $x>0$. We decompose the Brownian integral in (4.5) over the two sets corresponding to long and short paths: $\Delta_l(x)=\{b_x \vert 
\sup_{0\leq s\leq t}\vert b_x(s)\vert>{x\over 2}\}$ and 
$\Delta_s(x)=\{b_x \vert 
\sup_{0\leq s\leq t}\vert b_x(s)\vert<{x\over 2}\}$. If $b_x\in \Delta_s(x)$
we have $x+b_x(s)>{x\over 2}$ for all $s$ so that 
$U(x+b_x(s))\geq U({x\over 2})$ and
$$
\int_{\Delta_s(x)} Db_x e^{-2\int_0^t ds U(x+b_x(s))}
\leq e^{-2tU({x\over 2})}
\eqno(3.6)
$$
On the other hand if $b_x\in \Delta_l(x)$ we use $U(x +b_x(s))\geq 0$
and therefore
$$
\eqalign{\int_{\Delta_l(x)} Db_x e^{-2\int_0^t ds U(x+b_x(s))} &
\leq
\int_{\Delta_l(x)} Db_x
\cr &\leq 2\int_{b_x(t)\geq x}Db_x\leq 2\sqrt{2} e^{-{x^2\over 4t}}\cr}
\eqno(3.7)
$$
In (3.7) the second estimate is Levy's inequality (see [14, p 65]).
From (3.5), (3.6) and (3.7) we 
obtain for $x>0$
$$
\int_{-\infty}^{+\infty} dy\vert\Psi(\br)\vert^2
\leq 
{e^{2t(E+w_0)}\over (2\pi t)^{1/2}}\biggl(2\sqrt 2 e^{-{x^2\over 4t}}+
e^{-2tU({x\over 2})}\biggr)
\eqno(3.8)
$$
In (3.8) we are still free to choose $t>0$ as we wish. With 
$t=(x/2\sqrt 2)U(x/2)^{-1/2}$ the two exponential
terms in the bracket of (4.8) become equal
$$
\int_{-\infty}^{+\infty} dy\vert\Psi(\br)\vert^2
\leq 8{U({x\over 2})^{1/4}\over (2\pi x)^{1/2}}\exp\biggl((E+w_0)
{x\over\sqrt 2} U({x\over 2})^{-1/2}- {x\over \sqrt 2} U({x\over 2})^{1/2}\biggr)
\eqno(3.9)
$$
Let $x_0(U)$ be the solution of 
$$
U({x_0\over 2})=2(E+w_0)
\eqno(3.10)
$$
For 
$x>x_0(U)$ the estimate (4.9) becomes
$$
\int_{-\infty}^{+\infty} dy\vert\Psi(\br)\vert^2
\leq 8
{U({x\over 2})^{1/4}\over (2\pi x)^{1/2}}\exp\biggl(-{x\over 2\sqrt 2}
  U({x\over 2})^{1/2}\biggr)
\eqno(3.11)
$$
This bound does not depend on the magnetic field. It merely estimates the probability density of finding the 
quantum particle in the classically forbidden region by the wall.

\beginsection{4. ABSENCE OF EIGENVALUES IN-BETWEEN LANDAU LEVELS}

The proof of theorem 1 is based on the following two Lemmas.

\vskip 0.5cm

\noindent {\bf Lemma 1}

\noindent {\it Let $\Psi$ be a normalized eigenfunction of $H$, then
$$
\langle\Psi, (U^\prime+\partial_xw(x,y))\Psi\rangle=0
\eqno(4.1)
$$}
The relation (4.1) follows formally from (2.9) but requires a proof since $Y$ is
an unbounded function of the $y$ coordinate. The proof is given in Appendix C.

\vskip 0.5cm

\noindent {\bf Lemma 2}

\noindent {\it Let ${B\over 2}-w_0>\delta>0$ and $\Psi$ be a normalized 
eigenfunction of $H$ with eigenvalue 
$E\in \Delta_n(B,\delta)$. Then 
$$
\vert\vert U\Psi\vert\vert\geq {B\over 2}-\delta-w_0
\eqno(4.2)
$$
}

\vskip 0.5cm

\noindent {\bf Proof}

By hypothesis ${\rm dist}(E, \sigma(H(B))\geq {B\over 2}-\delta$,
therefore $(H(B)-E)^2\geq ({B\over 2}-\delta)^2$ which implies
$$
\vert\vert(H(B)-E)\Psi\vert\vert\geq 
({B\over 2}-\delta)\vert\vert\Psi\vert\vert={B\over 2}-\delta
\eqno(4.3)
$$
Moreover since $(H(B)-E)\Psi=-(U+w)\Psi$
 we have
$$
{B\over 2}-\delta\leq\vert\vert (U+w)\Psi\vert\vert\leq\vert\vert U\Psi\vert\vert+
\vert\vert w\Psi\vert\vert\leq\vert\vert U\Psi\vert\vert+w_0
\eqno(4.4)
$$
which gives the result (4.2).

\vskip 0.5cm

\noindent {\bf Proof of Theorem 1}

Suppose that $H$ has an eigenvalue $E\in \Delta_n(B,\delta)$ with corresponding 
normalized eigenstate $\Psi$.
From the Schwarz inequality
$$
\eqalign{
\vert\vert U\Psi\vert\vert^2 &=\int_0^{+\infty}dx\int_{-\infty}^{+\infty} dy
{U(x)^2\over U^\prime(x)^{1/2}}\vert\Psi(\br)\vert 
U^\prime(x)^{1/2}\vert\Psi(\br)\vert
\cr & \leq
\biggl\{\int_0^{+\infty}dx\int_{-\infty}^{+\infty} dy
{U(x)^4\over U^\prime(x)}\vert\Psi(\br)\vert^2\biggr\}^{1/2}
\biggl\{\int d\br U^\prime(x)\vert\Psi(\br)\vert^2\biggr\}^{1/2}
\cr}
\eqno(4.5)
$$
We decompose the $x$ integral in the right hand side of (5.7) into a part on
$0<x<x_0(U)$ and $x>x_0(U)$. Then using
$$
\int_0^{x_0(U)}dx\int_{-\infty}^{+\infty} dy
{U(x)^4\over U^\prime(x)}\vert\Psi(\br)\vert^2
\leq \sup_{0\leq x\leq x_0(U)}\biggl({U(x)^4\over U^\prime(x)}\biggr)
\eqno(4.6)
$$
and (3.11) for $x>x_0(U)$ we find
$$
\vert\vert U\Psi\vert\vert^4\leq A(E+w_0;U) \langle\Psi, U^\prime\Psi\rangle
\eqno(4.7)
$$
where  $A(E+w_0;U)$ is defined in (2.1). From Lemma 2 we obtain 
$$
\langle\Psi, U^\prime\Psi\rangle\geq 
{({B\over 2}-\delta-w_0)^4\over A(E+w_0;U)}
\eqno(4.8)
$$
On the other hand from Lemma 1 $\langle\Psi, U^\prime\Psi\rangle\leq w_0^\prime$, 
thus
$$
w_0^\prime\geq
{({B\over 2}-\delta-w_0)^4\over A(E+w_0;U)}
\eqno(4.9)
$$
Therefore $H$ cannot have eigenvalues in $\Delta_n(B,\delta)$ as long as (2.2)
holds.

\beginsection{5. EXISTENCE OF SPECTRUM}

The proof of theoreme 2 is based on the auxiliary Hamiltonian (5.1).
Consider a large disc of radius $R$, centered at the origin and take a smooth,
radially symmetric function $g_R(\br)$ satisfying
$g_R(\br)=1$ for $\vert\br\vert\leq R$, $g_R(\br)=0$ for $\vert\br\vert\geq R$,
$\vert\partial_ig_R(\br)\vert\leq 1/R$, $i=x,y$.
Set $w_R=-wg_R$ and 
$$
H_R=H+w_R=H_0+w(1-g_R)
\eqno(5.1)
$$
Note that by the results of Appendix A  $e^{-tH_R}-e^{-tH}$ is compact, hence
$H_R$ and $H$  have the same essential spectrum [13, Chap IV]. We show that 
in the neighbourhood of each energy $E$ in $\Delta_n(B,\delta)$ $H_R$, for $R$ 
large enough
(depending on $E$), has continuous spectrum. Hence 
because of the stability of the essential spectrum and the absence of eigenvalues,
$H$ will also have continuous spectrum around $E$.
\vskip 0.5cm

\noindent {\bf Lemma 4} 

\noindent{\it Under the assumptions of Theorem 2, for any $0<\delta^\prime<\delta$ there exist $R_0(\delta,\delta^\prime)$
such that for all $R>R_0(\delta,\delta^\prime)$, $H_R$ has no eigenvalue in 
$\Delta_n(B,\delta^\prime)$.}

\vskip 0.5cm

\noindent {\bf Proof}

Define $w_0(R)=\sup_\br\vert w(\br)(1-g_R(\br))\vert$ and 
$w_0^\prime(R)=\sup_\br\vert \partial_x(w(\br)(1-g_R(\br)))\vert$. We have
$w_0(R)\leq w_0$ and since ${B\over 2}-w_0>\delta$,
$$
{B\over 2}-w_0(R)>\delta^\prime
\eqno(5.2)
$$
Let us now check that (2.3) is satisfied for the Hamiltonian  $H_R$ and 
$\delta$ replaced by $\delta^\prime$. We have
$$
w_0^\prime(R)\leq \sup_\br\vert \partial_x w(\br)\vert\vert1-g_R(\br)\vert
+\sup_\br\vert w(\br)\partial_x(1-g_R(\br))\vert
\leq w_0^\prime+{w_0\over R}
\eqno(5.3)
$$
From (2.3) 
$$
\eqalign{w_0^\prime(R)&<{({B\over 2}-\delta-w_0)^4\over 
\sup_{0\leq w_0\leq{B\over 2}} \sup_{E\in \Delta_n(B,\delta)} A(E+w_0;U)}
+{w_0\over R}
\cr & =
{({B\over 2}-\delta^\prime-w_0)^4\over 
\sup_{0\leq w_0\leq{B\over 2}} \sup_{E\in \Delta_n(B,\delta)} A(E+w_0;U)}
+{w_0\over R}
\cr & +
{({B\over 2}-\delta-w_0)^4-({B\over 2}-\delta^\prime-w_0)^4\over 
\sup_{0\leq w_0\leq{B\over 2}} \sup_{E\in \Delta_n(B,\delta)} A(E+w_0;U)}
\cr}
\eqno(5.4)
$$
We see that the third term 
on the right hand side of the inequality is strictly negative
for $\delta^\prime<\delta$ and independent of $R$. So we can find 
$R_0(\delta,\delta^\prime)$ large enough
such that for all $R>R_0(\delta,\delta^\prime)$
$$
w_0^\prime(R)<{({B\over 2}-\delta^\prime-w_0)^4\over 
\sup_{0\leq w_0\leq{B\over 2}} \sup_{E\in \Delta_n(B,\delta)} A(E+w_0;U)}
\eqno(5.5)
$$
which implies the second inequality in (2.3) with $w_0$ replaced by
$w_0(R)$ and $\delta$ replaced by $\delta^\prime$. As a consequence of Theorem 1, $H_R$ has no eigenvalues in 
$\Delta_n(B,\delta^\prime)$
for $R>R_0(\delta,\delta^\prime)$. 

\vskip 0.5cm

\noindent {\bf Lemma 5}

\noindent {\it For any $E\in \Delta_n(B,\delta)$ and any $\epsilon>0$ we can 
find $R_1(\epsilon,E)$ such that for $R>R_1(\epsilon,E)$
$$
dist (E, \sigma(H_R))\leq\epsilon
\eqno(5.6)
$$
}

\vskip 0.5cm

\noindent{\bf Proof}

Since $\sigma(H_0)=[{B\over 2},\infty[$ we have $E\in\sigma(H_0)$. Thus there exist
$\Psi_0$, 
$$
\vert\vert(H_0-E)\Psi_0\vert\vert\leq {\epsilon\over 2}
\eqno(5.7)
$$
By the triangle inequality
$$
\eqalign{
\vert\vert(H_0+w(1-g_R)-E)\Psi_0\vert\vert & \leq {\epsilon\over 2}
+\vert\vert w(1-g_R)\Psi_0\vert\vert
\cr &
\leq {\epsilon\over 2}
+w_0\vert\vert(1-g_R)\Psi_0\vert\vert
\cr}
\eqno(5.8)
$$
Since $\vert\vert\Psi_0\vert\vert=1$ we can find $R_1(\epsilon,E)$ large enough such
that for $R>R_1(\epsilon,E)$ 
$$
\vert\vert(1-g_R)\Psi_0\vert\vert\leq {\epsilon\over
2w_0}
\eqno(5.9)
$$
From (5.8) and (5.9)
$\vert\vert(H_R-E)\Psi_0\vert\vert\leq \epsilon$,
and  (5.6) follows from the fact that for all $\Psi$,
$\vert\vert\Psi\vert\vert=1$, we have $dist(E,\sigma(H_R))\leq 
\vert\vert(H_R-E)\Psi\vert\vert$.

\vskip 0.5cm

\noindent{\bf Proof of Theorem 2}

Take $E\in \Delta_n(B,\delta^\prime)$. We suppose that $E$ belongs to 
the resolvent set of $H$ and show that it leads to a contradiction.
Since the resolvent is an open set we can find $\alpha>0$ such that
$dist(E,\sigma(H))\geq\alpha$. From the fact that $H_R$ and $H$ have the same 
essential spectrum we know that $H_R$ could have only isolated
eigenvalues in $]E-\alpha,E+\alpha[$, but this is impossible by Lemma 4 
for $R>R_0(\delta,\delta^\prime)$. Hence
$$
dist(E,\sigma(H_R))\geq\alpha, \qquad {\rm for}\qquad R>R_0(\delta,\delta^\prime)
\eqno(5.10)
$$
But if we take $\epsilon={\alpha\over 2}$ in Lemma 5 we have
$$
dist(E,\sigma(H_R))\leq {\alpha\over 2},\qquad {\rm for} 
\qquad R>R_1({\alpha\over 2},E)
\eqno(5.11)
$$
Therefore we obtain a contradiction for 
$R>\max\bigl(R_0(\delta,\delta^\prime),R_1({\alpha\over 2},E)\bigr)$. Thus
$E\in\sigma(H)$ for all $E\in \Delta_n(B,\delta^\prime)$.

\beginsection{6. CONCLUDING REMARKS}

We have shown that for a large class of wall potentials (those for 
which the integrals (2.1) and (B.8) are finite ) there are intervals of the
order $B-2w_0$ ($w_0$ small), centered  in-between Landau levels, where 
the spectrum is continuous. This does not prevent the possibility of having
point spectrum in the vicinity of the Landau levels.
In fact it is possible to construct an attractive, impurity potential with compact support which creates a bound state with energy 
${B\over 2}-w_0<E<{B\over 2}$ (recall that in this situation ${B\over 2}$ is the infimum of the continuous spectrum of 
$H_0$). In the random case one expects that a dense pure point spectrum will
form in the same energy interval, and that moreover for large
disorder the pure point spectrum extends also above ${B\over 2}$. 
In this connection see for example
[15] for general results on band edge localization. It would be interesting
to know whether this picture is valid or not near higher Landau levels.

A simpler situation is that of a homogeneous system without boundary and with crossed 
constant electric and magnetic fields, namely taking $U(x)=\mu x$ 
everywhere in space, with $\mu$ the amplitude of the electric field. 
Now the energy branches of 
$H_0$ are ${\cal E}_n(k)=(n+{1\over 2})B+{\mu\over B}k -{\mu^2\over 2B^2}$ 
so that $H_0$ has continuous spectrum on the whole  real line. 
Our previous analysis shows that the spectrum of $H=H_0+w$ remains continuous provided that
$\mu>w_0^\prime$. The absence of point spectrum of $H$ is now obvious for 
$\mu>w_0^\prime$ because of (2.10). The auxiliary Hamiltonian has no eigenvalue if
$\mu>w_0^\prime(R)$ (that is, in view of (6.3), for $R>w_0(\mu-w_0^\prime)^{-1}$), and 
has spectrum in the neighbourhood of any energy for $R$ large enough.
Then the result follows by the arguments used in theorem 2. 
An interesting question is the possible occurence of point spectrum for weak electric 
field ($\mu<w_0^\prime$). In the classical case, it is known that there exists
a set of trajectories with non zero Lebesgue measure that remain localized by the 
magnetic field around a hard obstacle, for non zero but sufficiently weak
electric field [16]. To our knowledge it has not been established wether
 this 
corresponds to a localized eigenstate in the quantum mechanical problem.

An other case of a system without boundaries, but now inhomogeneous, is a given
by a potential $U(x)$ that remains bounded as $x\to \infty$, for instance, 
the step potential
$$
U(x)=U_0(1-e^{-\alpha x}),\qquad x\geq 0,\alpha>0\qquad
{\rm and}\qquad U(x)=0,\qquad x\leq 0
$$
One can check that for $U_0$ large enough, theorems 1 and 2 apply, for energies in-between the lowest Landau levels, so that a
continuous spectrum will also occur in this system. If for example the energy 
is in the interval $]B-\delta, B+\delta[$ between the first and the second levels, 
sufficient conditions for the occurence of a continuous spectrum are $U_0\geq 4B$
(so that (3.10) has a solution) and $\sqrt{U_0}\geq 2\sqrt 2\alpha$
(ensuring that $A(E;U)$ in (2.1) is finite).

The notion of edge states could be made more precise by identifying them
to the subspaces of continuity corresponding to the intervals
$\Delta_n(B,\delta)$ determined in this work. This definition should 
be substantiated
by an analysis of the spatial behaviour of the generalized eigenfunctions of $H$ in 
these subspaces.

\beginsection{APPENDIX A: INTEGRABLE IMPURITY POTENTIALS}

Here we show that if $w$ is an integrable impurity potential, then
$e^{-tH}-e^{-tH_0}$ and $e^{-tH_R}-e^{-tH}$ are trace class.

The Feynman-Kac-Ito representation, together with the fact that $U(x)$ is positive and 
$w$ bounded leads to the bounds on the kernels
$$
\vert\langle\br_1\vert e^{-sH_0}\vert \br_2\rangle\vert\leq \langle 
\br_1\vert e^{{s\over 2}\Delta}\vert \br_2\rangle
\eqno(A.1)
$$
$$
\vert\langle\br_1\vert e^{-sH}\vert \br_2\rangle\vert\leq e^{sw_0}\langle
\br_1\vert e^{{s\over 2}\Delta}\vert \br_2\rangle
\eqno(A.2)
$$
for $0\leq s\leq t$, in terms of the free one 
$$
\langle\br_1\vert e^{{s\over 2}\Delta}\vert \br_2\rangle
={1\over 2\pi s}e^{-{\vert\br_1-\br_2\vert^2\over 2s}}
\eqno(A.3)
$$
If $\vert\vert...\vert\vert_{HS}$ denotes the Hilbert-Schmidt norm and $f(\br)$ is 
a square integrable function, a direct computation using (A.1), (A.2)
and (A.3) gives
$$
\vert\vert fe^{-sH_0}\vert\vert_{HS}\leq {\vert\vert f\vert\vert_2\over \sqrt{4\pi s}},\qquad
\vert\vert fe^{-sH}\vert\vert_{HS}\leq e^{tw_0}{\vert\vert f\vert\vert_2\over \sqrt{4\pi s}}
\eqno(A.4)
$$
The trace norm of $e^{-tH}-e^{-tH_0}$ is majorized by
$$
\eqalign{\vert\vert e^{-tH}-e^{-tH_0}\vert\vert_1 &
\leq
\int_0^t ds \vert\vert e^{-(t-s)H_0}we^{-sH}\vert\vert_1
\cr &
\leq 
\int_0^t ds \vert\vert e^{-(t-s)H_0}\sqrt{\vert w\vert}\vert\vert_{HS}
\vert\vert\sqrt{\vert w\vert}e^{-sH}\vert\vert_{HS}
\cr}
\eqno(A.5)
$$
and thus is finite by (A.4) when $w$ is integrable. The proof is the same
for $e^{-tH_R}-e^{-tH}$.

\beginsection{APPENDIX B: SPECTRUM OF $H_\omega$}

We show that $[{B\over 2},\infty[\subset\sigma(H_\omega)\subset [{B\over 2}-w_0,\infty[$. 
The second inclusion simply follows from 
$infspec H_{\omega}\geq {B\over 2}-w_0$, for all $\omega$.

For the first inclusion it is sufficient to prove that, given
$E\in [{B\over 2},\infty[$, for any $\epsilon>0$ there exist $\Omega$ with
$Prob(\Omega)>0$ and $\Psi$, $\vert\vert\Psi\vert\vert=1$, such that
$\vert\vert (H_\omega-E)\Psi\vert\vert\leq \epsilon$ 
for all $\omega\in \Omega$.

Since $\sigma(H_0)=[{B\over 2}, \infty[$, given any $E\geq {B\over 2}$ and
any $\epsilon>0$, there exist a $\Psi$, $\vert\vert\Psi\vert\vert=1$,
such that
$$
\vert\vert (H_0-E)\Psi\vert\vert\leq {\epsilon\over 3}
\eqno(B.1)
$$
Take $\Omega=\{\omega \vert \vert\omega_{nm}\vert\leq{\epsilon\over 3\sup_\br
\vert v(\br)\vert}, {\rm  for} (n,m)\in B_L\}$, where $B_L$ is a 
square of size $2L+1$ centered at the origin. Note that $Prob(\Omega)>0$ 
and for all $\omega\in \Omega$
$$
\vert\vert V_\omega(\br)\Psi\vert\vert\leq
{\epsilon\over 3} + \sup_\br\vert v(\br)\vert \vert\vert \chi_{B_L^c}\Psi\vert\vert
\eqno(B.2)
$$
where $\chi_{B_L^c}$ is the characteristic function of the complement of
$B_L$. Taking $L$ large enough so that the last term on the right hand side
of this inequality is less than ${\epsilon\over 3}$ we obtain
$$
\vert\vert (H_0-E)\Psi\vert\vert+
\vert\vert V_\omega(\br)\Psi\vert\vert\leq\epsilon
\eqno(B.3)
$$
Thus by the triangle inequality 
$\vert\vert (H_\omega-E)\Psi\vert\vert\leq \epsilon$.

\beginsection{APPENDIX C: PROOF OF LEMMA 1}

Set $D=U^\prime+\partial_x w$. The estimate (3.11) with (1.1) implies 
$\vert\vert D\Psi\vert\vert<\infty$. defining $R_\lambda=\lambda(iY+\lambda)^{-1}$,
$\lambda>0$, we will show below that also
$$
\vert\vert DR_\lambda\Psi\vert\vert<\infty
\eqno(C.1)
$$
and 
$$
\vert\vert HR_\lambda\Psi\vert\vert<\infty
\eqno(C.2)
$$
uniformly with respect to $\lambda>0$.

Note that from (C.2), $HYR_\lambda\Psi=-i\lambda H(R_\lambda-I)\Psi$ has a finite
norm. Then the following  identities hold
$$
D\Psi=-(R_\lambda-I) D(R_\lambda-I)\Psi+
(R_\lambda-I) D\Psi+D(R_\lambda-I)\Psi +R_\lambda DR_\lambda\Psi
\eqno(C.3)
$$
$$
R_\lambda DR_\lambda\Psi=i(R_\lambda YHR_\lambda\Psi-R_\lambda HYR_\lambda\Psi)
=\lambda(HR_\lambda\Psi-R_\lambda H\Psi)
\eqno(C.4)
$$
From (C.4) we find $\langle\Psi,R_\lambda DR_\lambda\Psi\rangle=0$ 
since $\Psi$ is an eigenvector of $H$. Using (C.3) we 
conclude that $\langle\Psi, D\Psi\rangle=0$ by letting $\lambda\to \infty$ and noting
that $R_\lambda-I$ tends strongly to zero.

To show (C.1) and (C.2), we use the fact that $e^{iYa}=e^{iBya}e^{-ip_xa}$ is 
the operator of translations in the $x$ direction (up to a phase). This 
leads to the formula
$$
(R_\lambda\Psi)(x,y)=\lambda\int_0^\infty da e^{-\lambda a}e^{iBya}\Psi(x+a, y)
\eqno(C.5)
$$
Splitting $e^{-\lambda a}=e^{-\lambda {a\over 2}}e^{-\lambda {a\over 2}}$
we obtain by an application of the Schwartz inequality
$$
\eqalign{\int_{-\infty}^{+\infty} dy\vert 
(R_\lambda\Psi)(x,y)\vert^2
 &
\leq\lambda\int_0^\infty da e^{-\lambda a} 
\int_{-\infty}^{+\infty} dy\vert\Psi(x+a,y)\vert^2
\cr &
\leq C\exp\biggl[-{x\over 4\sqrt 2}U({x\over 2})^{1/2}\biggr]
\cr}
\eqno(C.6)
$$
The second inequality holds for  all $x\geq x_0>0$. It follows from (3.11) and the 
monotonicity of $U(x)$ with $C$ independent of $\lambda$.
Hence, noting that $\vert\vert R_\lambda\vert\vert=1$, this provides the
uniform upper bound with respect to $\lambda$
$$
\vert\vert U^\prime R_\lambda\Psi\vert\vert^2\leq\sup_{0\leq x\leq x_0}U^\prime(x)
+C
\int_{x_0}^{+\infty}
U^\prime(x)^2\exp\biggl[-{x\over 4\sqrt 2}U({x\over 2})^{1/2}\biggr]
\eqno(C.7)
$$
The same is also true for $\vert\vert DR_\lambda\Psi\vert\vert^2$ since $\partial_x w$
is bounded, so (C.1) holds. Since $Y$ commutes with $H(B)$ one has
$$
HR_\lambda\Psi=ER_\lambda\Psi+[U,R_\lambda]\Psi+[w,R_\lambda]\Psi
\eqno(C.8)
$$
Using estimate (1.1) and (C.6) we find $\vert\vert U\Psi\vert\vert<\infty$
and $\vert\vert UR_\lambda\Psi\vert\vert<\infty$, so that all 
vectors in the right hand side of (C.8) have finite norm, hence (C.2) holds.

\beginsection{REFERENCES}

\noindent [1] B. I. Halperin, Phys. Rev. B {\bf 25}, 2185 (1982)

\noindent [2] X. G. Wen, Phys. Rev. B {\bf 43}, 11025 (1991)

\noindent [3] J. Froehlich, T. Kerler, Nucl. Phys. B {\bf 354}, 369 (1991)

\noindent [4] for a review with a complete list of references see
J. Froehlich, U. M. Studer, Rev. Mod. Phys {\bf 65}, 733 (1993)

\noindent [5] A. M. Chang, L. N. Pfeiffer, K. W. West, Phys. Rev. Lett {\bf 77}, 2538 (1996); see also Physics Today {\bf 49}, p 19 (September 1996)

\noindent [6] T. C. Dorlas, N. Macris, J. V. Pul\'e, J. Stat. Phys {\bf 87}, 847
(1997); "Characterization of the spectrum of the Landau hamiltonian with delta 
impurities" preprint (1998); "Quantum Hall effect without divergence of the 
localization length" preprint (1998)

\noindent [7] T. C. Dorlas, N. Macris, J. V. Pul\'e, Helv. Phys. Acta {\bf 68}, 330 (1995); J. Math. Phys {\bf 37}, 1574 (1996) 

\noindent [8] J. M. Combes, P. D. Hislop, Comm. Math. Phys 
{\bf 177}, 603 (1996)

\noindent [9] W. M. Wang, J. Funct. Anal {\bf 146}, 1 (1997)

\noindent [10] J. Froehlich, private communication, and G. M. Graf, J. Walcher, "Anwendung
positiver kommutatoren auf den quanten Hall effect", Diploma work ETHZ (1998)

\noindent [11] M. Reed, B. Simon, "Methods of modern mathematical physics",
Volume 2, Academic Press, New York (1978)

\noindent [12] M. Reed, B. Simon, "Methoda of modern mathematical physics",
 Volume 4, 
Academic Press, New York (1978)

\noindent [13] T. Kato, "Perturbation theory for linear operators", second edition,
Springer Verlag, Berlin, Heidelberg, New York (1980)

\noindent [14] B. Simon, "Functional integration and quantum physics", Academic Press, New York (1979) 

\noindent [15] J. M. Barbaroux, J. M. Combes, P. D. Hislop, Helv. Phys. Acta
{\bf 70}, 16 (1997)

\noindent [16] N. Berglund, A. Hansen, E. H. Hauge, J. Piasecki, Phys. Rev. 
Lett {\bf 77} 2149 (1996)

\bye